\documentclass[12pt]{article}
\usepackage{graphicx}

\newcommand\pubnumber{SNSN-323-63}
\newcommand\pubdate{\today}

\def\osaka{Department of Physics, Osaka University\\
Toyonaka, Osaka 560-0043, Japan}

\def\Title#1{\begin{center} {\Large #1 } \end{center}}
\def\Author#1{\begin{center}{ \sc #1} \end{center}}
\def\Address#1{\begin{center}{ \it #1} \end{center}}

\newcommand\pubblock{\rightline{\begin{tabular}{l} \pubnumber\\
         \pubdate  \end{tabular}}}
\newenvironment{Abstract}{\begin{quotation}  }{\end{quotation}}
\newenvironment{Presented}{\begin{quotation} \begin{center} 
             PRESENTED AT\end{center}\bigskip 
      \begin{center}\begin{large}}{\end{large}\end{center} \end{quotation}}

\def\beq{\begin{equation}}
\def\eeq#1{\label{#1}\end{equation}}
\def\eeqn{\end{equation}}

\def\beqa{\begin{eqnarray}}
\def\eeqa#1{\label{#1}\end{eqnarray}}
\def\eeqan{\end{eqnarray}}

\let\bar=\overbar

\def\Dslash{\not{\hbox{\kern-4pt $D$}}}
\def\dslash{\not{\hbox{\kern-2pt $\del$}}}

\def\msb{{\bar{\ssstyle M \kern -1pt S}}}

\newcommand{\kpnn}{$K^{0}_{L}\rightarrow \pi^{0} \nu \bar{\nu}$}

\newcommand{\kpitwo}{$K^{0}_{L}\rightarrow 2 \pi^{0}$}
\newcommand{\kpithree}{$K^{0}_{L}\rightarrow 3 \pi^{0}$}
\newcommand{\ktwogamma}{$K^{0}_{L}\rightarrow 2{\gamma}$}
\newcommand{\kpithreec}{$K^{0}_{L}\rightarrow \pi^{+} \pi^{-} \pi^{0}$}
\newcommand{\kpienu}{$K^{0}_{L}\rightarrow \pi^{\pm}e^{\mp}\nu$}
\newcommand{\kpimunu}{$K^{0}_{L}\rightarrow \pi^{\pm}\mu^{\mp}\nu$}

\newcommand{\pklong}{$K^{0}_{L}$}
\newcommand{\ppi}{$\pi^{0}$}

\begin{document}
\begin{titlepage}
\pubblock

\vfill
\Title{ {\kpnn} at KOTO }
\vfill
\Author{ Koji Shiomi for the KOTO collaboration}
\Address{\osaka}
\vfill
\begin{Abstract}

The KOTO experiment aims to discover the rare decay {\kpnn} at J-PARC.
This mode breaks the CP symmetry directly and is highly suppressed 
in the Standard Model.
Thus the mode is sensitive to new physics,
in particular to the models related to CP violation.

We performed the first physics run in May 2013.
Although the data taking was terminated after 100 hours due to an accident of the facility,
the single event sensitivity for the {\kpnn} decay reached $1.29\times10^{-8}$,
which was comparable to the sensitivity of the prior experiment.
We observed one candidate event while
$0.36\pm0.16$ background events were expected.

\end{Abstract}
\vfill
\begin{Presented}
The 8th International Workshop on the CKM Unitarity Triangle (CKM 2014) \\
Vienna, Austria, September 8--12, 2014
\end{Presented}
\vfill
\end{titlepage} 
\def\thefootnote{\fnsymbol{footnote}}
\setcounter{footnote}{0}

\section{Introduction}
The neutral-kaon decay {\kpnn} breakes the CP symmetry directly.
The decay is highly suppressed in the standard model (SM)
because it is a Flavor-Changing Neutral Current (FCNC) process, 
which is induced through electroweak loop diagrams.
The branching ratio of the {\kpnn} decay is predicted to be 
$2.4\times10^{-11}$ in the SM and the theoretical uncertainty 
is only a few percent\cite{BRkpnn}.  
Owing to these features, the decay is sensitive to new physics contributions 
that may appear in loop diagrams.
Several theoretical models of new physics, which 
introduce new CP-violating phases, predict higher branching ratios
than that of the SM prediction\cite{MSSM,4Gen}. 
The current best upper limit on the decay is $2.6\times10^{-8}$ at
the 90\% confidence level, which was set by the KEK E391a 
experiment\cite{E391a}. 

The KOTO experiment is an upgrade of the E391a experiment
and aims at the first observation of the decay at the J-PARC Main Ring accelerator.
To achieve this goal, we have built a new high-intensity neutral-kaon beam line and
constructed the new detector system and DAQ system.
We performed engineering and commissioning runs to study
the detector performance in early 2013,
and started the first physics run in May 2013 with the 24 kW beam
which corresponds to 10\% of the designed power.
However, the data taking was terminated after 100 hours due to an accident of the facility.

In this manuscript, preliminary results from the first physics run are reported.

\section{Experiment}
A 30-GeV proton beam was extracted to the hall of J-PARC Hadron Experimental Facility
every 6 seconds with a 2-second flat top, and 
was injected into a 66-mm-long gold target.
{\pklong}'s produced at 16$^\circ$ from the proton beam were transported to the detector system
through a 21-m-long beam line consisting of two collimators, a sweeping magnet, and a lead photon absorber.

\begin{figure} [htbp]
 \centering
  \includegraphics[width=15cm,clip]{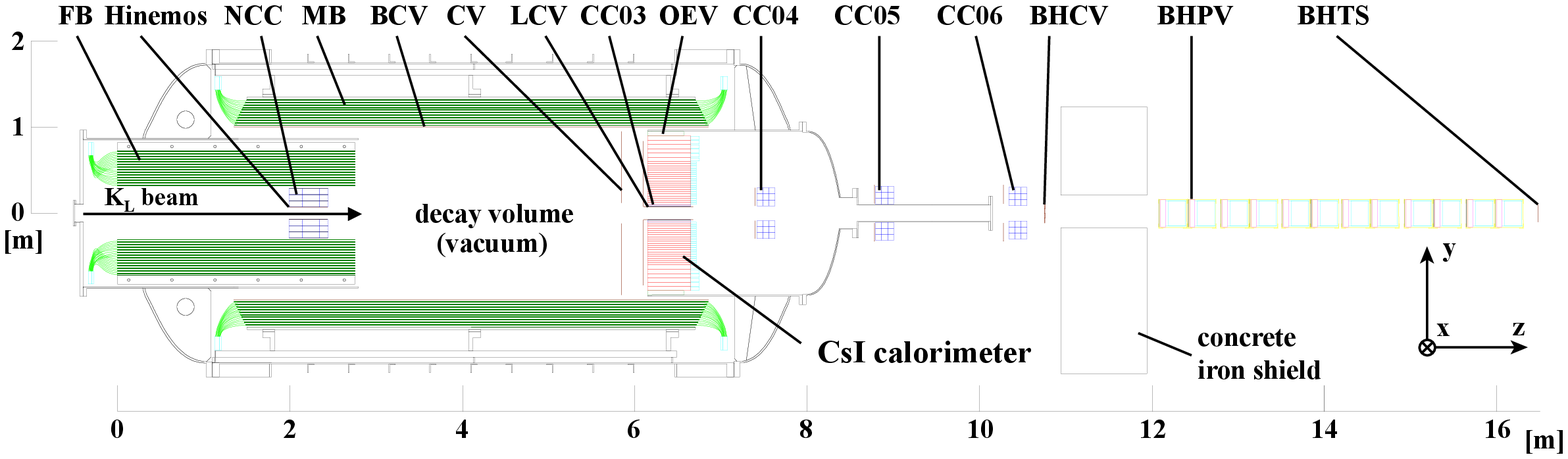}
 \caption{Cross-sectional side view of the KOTO detector system. The {\pklong} beam comes in 
 from the left hand side. Also shown is the coordinate system used in this paper: 
 $x$ is the horizontal, $y$ is the vertical, and $z$ is the beam directions.
 The origin of the coordinate is set at the upstream end of FB.}
 \label{fig:KOTOdet}
\end{figure}

The signature of the {\kpnn} decay is two photons from the {\ppi} decay and no other
particles in the final state.
The KOTO detector consists of an electromagnetic calorimeter 
to measure two photons from {\ppi}, and veto detectors surrounding the decay region 
to detect all particles not hitting the calorimeter, as shown in Fig.~\ref{fig:KOTOdet}.

The electromagnetic calorimeter is made of 2716 undoped CsI crystals.
The front of the calorimeter is covered with a set of plastic scintillators,
named "Charged Veto(CV)".
The decay region is surrounded with two large lead-scintillator sandwich counters,
called "Main Barrel (MB)" and "Front Barrel (FB)".
Inside FB, a veto detector called "Neutron Collaor Counter (NCC)" is located
to prevent particles from  the upstream region of the decay region from hitting the calorimeter directly.
A series of veto detectors called "Collaor Counters (CCs)" are placed 
to detect particles that go through the beam hole of the calorimeter.
At the downstream end of the detector system, "Beam Hole" veto detectors 
are located in the beam.

\section{Analysis}

\begin{figure} [htbp]
 \centering
 \includegraphics[width=9cm]{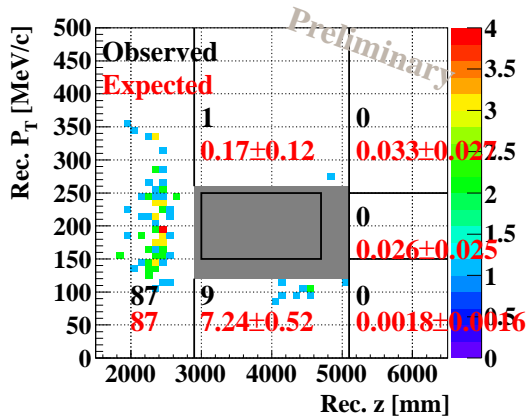}
 \caption{Scatter plot of Rec. P$_{T}$ vs Rec. z after imposing all selection criteria.
 The dark box is the masked region in the analysis. The inner box is the signal region.
 The numbers in black show the number of observed events, and
 the numbers in red show the number of expected background events based on Monte Carlo simulation.}

 \label{fig:PtZ_w_mask}
\end{figure}

In the KOTO experiment, the energies and positions of the two photons 
were measured with the CsI calorimeter.
The decay position of the {\ppi} along the beam axis was reconstructed,
assuming that the invariant mass of the two photons had the {\ppi} mass.
Using the decay vertex, the transverse momentum of the {\ppi} was also calculated.
We required a finite transverse momentum to select signal events because 
neutrinos should have carried away a finite transverse momentum. 
In addition,  we required that there were no other activities in hermetic veto counters
surrounding the decay region to confirm that only one {\ppi} existed in the final state.
The kinematical selection criteria were applied to distinguish signal events from background events.
The signal region for the {\kpnn} decay was defined in the scatter plot of the reconstructed vertex
position (Rec.~z) and the transverse momentum of {\ppi} (Rec.~P$_{T}$)  
to be $3000\leq {\rm Rec.~z} \leq 4700$ mm and $150 \leq {\rm Rec.~P}_{T} \leq 250$ MeV/c, respectively.
Figure \ref{fig:PtZ_w_mask} shows the plot in the Rec.~P$_{T}$ vs Rec.~z after imposing all selection criteria.

There is a cluster of events located at Rec.~z = $2000-2600$  mm.
These events were {\ppi}'s produced by neutrons in the halo of the beam (halo neutrons)
interacting in NCC. 
Because the two photons from the {\ppi} hit the CsI calorimeter,
the decay vertex was reconstructed at the NCC location that was outside the signal region in Rec.~z.
We call this type of background events "upstream events".
The number of upstream events in the signal region was estimated to be $0.06\pm0.06$ based on 
the Monte Carlo simulation (MC).

The events below the signal region came from the {\kpithreec} decay.
In these events, two charged pions from {\kpithreec} escaped through 
the beam hole in the calorimeter and interacted with the vacuum pipe along the beam axis.
These events could also hardly entered the signal region 
because we required the transverse momentum of {\ppi} to be larger than
the kinematical limit for {\kpithreec}.
The number of {\kpithreec} events in the signal region was estimated to be less than 0.01 based on MC.

The event in Fig.~\ref{fig:PtZ_w_mask} in the region above the signal box 
was likely to be caused by halo neutrons.
We suspect that a single neutron hitting the CsI calorimeter 
could generate two clusters through several hadronic interactions in the calorimeter.
We call this type of background events "hadron interaction events",
which were the most serious background source in the first physics run.
We used special data to study the background.
In this data, a 5-mm-thick aluminum disk was inserted into the beam
to scatter neutrons into the CsI calorimeter.
By using the special data, the number of hadron interaction events inside the 
signal region was estimated to be $0.18\pm0.15$.

Table \ref{tab:BG} summarizes the number of background events expected in the signal region.
In addition to the background sources described above, 
we estimated the backgrounds originated by the {\kpienu}, {\kpimunu}, {\kpithree}, {\ktwogamma}, and {\kpitwo}  decays.
These events were suppressed to be less than 0.1 for each decay mode. 
The sum of the {\pklong} background expectations was $0.11\pm0.04$.
The total number of background events in the signal region was estimated to be $0.36\pm0.16$.

\begin{table}[t]
\begin{center}
\begin{tabular}{|c|c|}  \hline
Background source             &  The number of background events \\ \hline \hline
Hadron interaction events & $0.18\pm0.15$ \\ \hline
Kaon decay events             & $0.11\pm0.04$ \\ \hline
Upstream events                 & $0.06\pm0.06$ \\ \hline \hline
 Sum                                      & $0.36\pm0.16$ \\ \hline
\end{tabular}
\end{center}
\caption{Estimated numbers of background events in the signal region.}
\label{tab:BG}
\end{table}

\section{Results}
\begin{figure} [htbp]
 \centering
 \includegraphics[width=9cm]{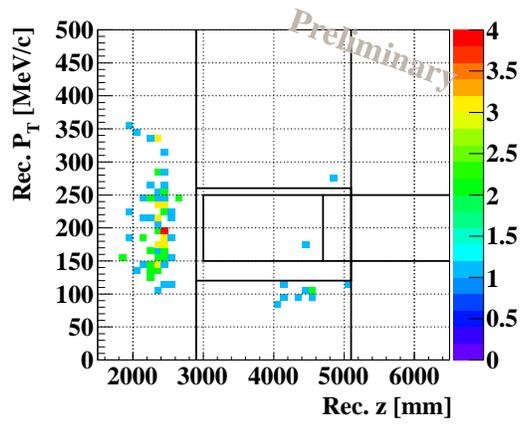}
 \caption{Scatter plot in the Rec. P$_{T}$ vs Rec. z after imposing all selection criteria.
 The inner box is the signal region.}
 \label{fig:PtZ}
\end{figure}

After determining all the selection criteria and estimating background levels,
we examined the events in the signal region and observed one candidate,
as shown in Fig.~\ref{fig:PtZ},
which was statistically consistent with the number of expected background events.

Based on the number of the simultaneouly-measured {\kpitwo} events, 
we evaluated that the single event sensitivity (S.E.S) for the {\kpnn} decay 
was $1.29\times10^{-8}$, while the
S.E.S of the E391a experiment was $1.11\times10^{-8}$.
Although data taking time was only 100 hours we achieved
the same sensitivity as E391a, which took data for 1100 hours.

\section{Conclusion}
The KOTO experiment, the {\kpnn} study at J-PARC,
took the first physics data in May 2013.
Though the run was only 100-hour long, 
we achieved the same sensitivity as the KEK E391a experiment.
One event was observed, which
was statistically consistent with the number of expected background events.

We are preparing for the next data taking,
which we plan to start in early 2015. 
To improve the experimental sensitivity, we are preparing new
detectors to reduce the background events and are improving 
analysis methods to remove hadron interaction events.


\end{document}